\documentclass[aps,preprint,showpacs,groupedaddress]{revtex4-1}
\usepackage{amssymb}
\usepackage{mathrsfs}
\usepackage{amsmath}
\usepackage{mathtools}
\usepackage{pstricks}
\usepackage{color}
\usepackage{graphicx}
\usepackage{amsthm}

\usepackage{physics}

\begin{document}


\title{\bf Influence of conformal symmetry on the amplitude ratios for O($N$) models}



\author{H. A. S. Costa}
\email{hascosta@ufpi.edu.br}
\affiliation{\it Departamento de F\'\i sica, Universidade Federal do Piau\'\i, 64049-550, Teresina, PI, Brazil}

\author{P. R. S. Carvalho}
\email{prscarvalho@ufpi.edu.br}
\affiliation{\it Departamento de F\'\i sica, Universidade Federal do Piau\'\i, 64049-550, Teresina, PI, Brazil}




\begin{abstract}
In this Letter we compute analytically the effect of conformal symmetry on the radiative corrections to the amplitude ratios for O($N$) $\lambda\phi^{4}$ massless scalar field theories in curved spacetime for probing the two-scale-factor universality hypothesis. For that we employ three distinct and independent field-theoretic renormalization group methods. The amplitude ratios values obtained are identical when computed through the three distinct methods, thus showing their universal character. Furthermore, they are the same as that obtained in flat spacetime, then satisfying the two-scale-factor universality hypothesis.  
\end{abstract}


\maketitle


\section{Introduction}\label{Introduction} 

\par Symmetry is one of the most important properties to consider if we want to describe the behavior of physical systems. For example in the high energy physics scenario, the standard model (SM) of elementary particles and fields which describes three of the four elementary interactions of nature, namely electromagnetic, weak and strong interactions, is based on gauge symmetries \cite{Itzykson,Peskin}. The remaining elementary interaction, which is not part of the SM, \emph{i. e.} gravitation \cite{Misner.Thorne.Wheeler,Wald} is also a gauge theory. On the other hand, in the low energy physics realm the importance of symmetry considerations is not quite different as well. In fact, the universal critical scaling behavior based on the scaling hypothesis \cite{doi10106311696618,PhysicsPhysiqueFizika.2.263} of systems undergoing a continuous phase transition is characterized by the determination of a set of universal critical exponents. These universal critical exponents do not depend on nonuniversal properties of the system as the form of the lattice and critical temperature but on universal ones as the dimension $d$, $N$ and symmetry of some $N$-component order parameter and if the interactions among their constituents are of short- or long-range type. Obviously the $d$ \cite{PhysRevB.86.155112,PhysRevE.71.046112} and $N$ \cite{PhysRevLett.110.141601,Butti2005527,PhysRevB.54.7177} parameters are more evident to study as opposed to the symmetry one \cite{CARVALHO2017290,Carvalho2017}. Surprisingly, the critical behavior of many distinct systems as a fluid and a ferromagnet is characterized by the same set of critical exponents. When this happens we say that the different systems belong to the same universality class. The universality class approached here will be the O($N$) one which generalizes and encompass some particular models, namely the Ising ($N=1$), XY ($N=2$), Heisenberg ($N=3$), self-avoiding random walk ($N=0$), spherical ($N \rightarrow \infty$) \cite{Pelissetto2002549} ones. The critical exponents are not the only universal quantities in describing the critical behavior of systems undergoing a phase transition. This role is similarly played by others universal quantities, although being harder to compute than the former ones, called amplitude ratios \cite{V.PrivmanP.C.Hohenberg}. Then emerges the concept of two-scale-factor universality \cite{PhysRevLett.29.345}, where now there are eleven independent universal amplitude ratios since we have chosen two independent length scales as the order parameter and conjugate field scales for example. While the critical exponents are universal quantities, the amplitudes themselves are not since they depend on nonuniversal properties of the system. The universal quantities are, in fact, some ratios involving some non universal amplitudes, where the nonuniversal properties, explicitly expressed in the amplitudes themselves, cancel out in the middle of computations and thus turn out the amplitude ratios to be universal quantities. 

\par In this Letter we have to investigate the influence of conformal symmetry on the values of the amplitude ratios for massless O($N$) $\lambda\phi^{4}$ scalar field theory in curved spacetime. When we try to renormalize a massless theory in curved spacetime following the conventional program, some divergences yet persist \cite{0305-4470-13-3-023}. The fixed value of $\xi = 1/6$ defines the theory as being invariant under conformal transformations. The amplitude ratios are a result of the fluctuating properties of a fluctuating scalar quantum field $\phi$ whose mean value we can identify to the order parameter, the magnetization of the system below the critical temperature $T_{c}$ for example. Since the theory involves properties of the system at the low temperature phase, as the magnetization, we have to describe a theory with spontaneous symmetry breaking since some amplitude ratios involve a few critical amplitudes computed below the critical temperature. Another feature showed through the lower temperature phase is that due to the presence of Goldstone modes, we have that some amplitudes and thus some amplitude ratios are not defined for every $N$ but only for Ising-like systems for which $N = 1$. These are the cases of both $C^{-}$ and $\xi_{0}^{-}$ amplitudes. The effect of quantum field fluctuations is evaluated as coming from the radiative quantum corrections to the renormalized effective potential with spontaneous symmetry breaking. If we do not take these loop corrections into account, we are limited to obtain the amplitude ratios values associated to the mean field or Landau approximation \cite{ZinnJustin}. We compute the amplitude ratios up to one-loop order. The effective potential with spontaneous symmetry breaking is renormalized in the normalization conditions method \cite{Amit}, minimal subtraction scheme \cite{Amit} and massless Bogoliubov-Parasyuk-Hepp-Zimmermann (BPHZ) method \cite{BogoliubovParasyuk,Hepp,Zimmermann}, where the external momenta of Feynman diagrams are held at fixed values in the first method and arbitrary in the last two ones. Although the application of a single method in computing the amplitude ratios is enough, the application of two another ones, besides to be useful as a check of the final results, must furnish the same amplitude ratios values as the renormalization group program demands since these physical quantities are universal. The same task was approached in a flat spacetime with Lorentz symmetry breaking mechanism \cite{Neto2017}. We then follow the steps of Ref. \cite{Neto2017} taken originally in Ref. \cite{BrezinLeGuillouZinnJustin}. In this Letter, the fluctuating quantum field is embedded on a curved spacetime and considering its non-minimal interaction with the curved background of the form $\xi R\phi^{2}$, where $\xi$ and $R$ are the non-minimal interaction coupling constant and the scalar curvature $R = g^{\mu\nu}R_{\mu\nu}$, respectively \cite{PhysRevD.20.2499,0305-4470-13-3-022,0305-4470-13-2-023,Bunch1981,Buchbinder1989,Class.QuantumGrav.}. We have to expand the free propagator of the theory and to make our calculations up to linear powers in $R$ and $R_{\mu\nu}$. Also up to linear powers in $R$ and $R_{\mu\nu}$, the present authors performed a computation of the critical exponents in an early work \cite{Costa_2019}. Now we proceed to obtain the corresponding amplitude ratios. 

\section{Normalization conditions method}\label{Normalization conditions method}

\par The critical amplitudes \cite{Neto2017}
\begin{eqnarray}\label{uhdfuhfuh}
\begin{array}{lcr}
\mbox{\textrm{Critical isochore: $T > T_{c}$, $H = 0$}} &  &  \\
\mbox{} \xi = \xi_{0}^{+}t^{-\nu}, \chi = C^{+}t^{-\gamma}, C_{s} = \frac{A^{+}}{\alpha_{+}}t^{-\alpha} &  \\ [10pt]
\mbox{\textrm{Critical isochore: $T < T_{c}$, $H = 0$}} &  &  \\
\mbox{} \xi = \xi_{0}^{-}t^{-\nu}, \chi = C^{-}t^{-\gamma}, C_{s} = \frac{A^{-}}{\alpha_{-}}t^{-\alpha}, M = B(-t)^{\beta} &  \\ [10pt]
\mbox{\textrm{Critical isotherm: $T = T_{c}$, $H \neq 0$}} &  &  \\ 
\mbox{} \xi = \xi_{0}^{c}|H|^{-\nu_{c}}, \chi = C^{c}|H|^{-\gamma_{c}}, C_{s} = \frac{A^{c}}{\alpha_{c}}|H|^{-\alpha_{c}}, H = DM^{\delta} &  \\ [10pt]
\mbox{\textrm{Critical point: $T = T_{c}$, $H = 0$}} &  &  \\
\mbox{} \chi(p) = \widehat{D}p^{\eta-2} &  \\ \end{array} \nonumber
\end{eqnarray}
are obtained through the renormalized effective potential with spontaneous symmetry breaking, equation of state and the longitudinal and transverse correlation functions, respectively \cite{Neto2017}
\begin{eqnarray}\label{fhdldglkjdflk}
&& \mathcal{F}(t,M,g^{\ast}) =  \frac{1}{2}tM^{2} + \frac{1}{4!}g^{\ast}M^{4} +  \frac{1}{4}\left[ Nt^{2} + \frac{N+2}{3}tg^{\ast}M^{2} + \frac{N+8}{36}(g^{\ast}M^{2})^{2} \right]\parbox{10mm}{\includegraphics[scale=1.0]{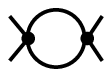}}_{SP} + \nonumber \\ &&  \frac{1}{2}\int d^{d}q\Bigg\{ \ln\Bigg[ 1 + \Bigg(t + \frac{g^{\ast}M^{2}}{2}\Bigg)G_{0}(q) \Bigg] +  (N-1)\ln\Bigg[ 1 + \Bigg(t + \frac{g^{\ast}M^{2}}{6}\Bigg)G_{0}(q) \Bigg]  - \nonumber \\ && \frac{N+2}{6}tg^{\ast}M^{2}G_{0}(q) \Bigg\},
\end{eqnarray}
\begin{eqnarray}\label{jglkfjkflj}
&& H/M = t + \frac{1}{6}g^{\ast}M^{2} +  \frac{1}{2}g^{\ast}\Biggl\{ \Biggl[ \parbox{12mm}{\includegraphics[scale=1.0]{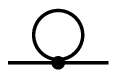}}_{(1)} + (t + g^{\ast}M^{2}/2)\parbox{10mm}{\includegraphics[scale=1.0]{fig10.eps}}_{SP} \Biggl]  + \nonumber \\ &&  \frac{N - 1}{3}\Biggl[ \parbox{12mm}{\includegraphics[scale=1.0]{fig1.eps}}_{(N - 1)} + (t + g^{\ast}M^{2}/6)\parbox{10mm}{\includegraphics[scale=1.0]{fig10.eps}}_{SP} \Biggl] \Biggl\},\nonumber \\ 
\end{eqnarray}
\begin{eqnarray}\label{uhduhuhg}
&& \Gamma_{L}(P^{2},t,M) = P^{2} + \frac{1}{6}R_{\mu\nu}\frac{\partial}{\partial P_{\mu}}\frac{P^{\nu}}{(P^{2} + t + g^{\ast}M^{2}/2)^{2}}  + t + \frac{1}{2}g^{\ast}M^{2} + \nonumber \\ && \frac{1}{2}g^{\ast}\Biggl\{ \Biggl[ \parbox{12mm}{\includegraphics[scale=1.0]{fig1.eps}}_{(1)} + (t + g^{\ast}M^{2}/2)\parbox{10mm}{\includegraphics[scale=1.0]{fig10.eps}}_{SP} \Biggl]  + \frac{N - 1}{3}\Biggl[ \parbox{12mm}{\includegraphics[scale=1.0]{fig1.eps}}_{(N - 1)} + (t + g^{\ast}M^{2}/6)\parbox{10mm}{\includegraphics[scale=1.0]{fig10.eps}}_{SP} \Biggl] + \nonumber \\ && g^{\ast}M^{2}\Biggl[ \Biggl(\parbox{10mm}{\includegraphics[scale=1.0]{fig10.eps}}_{SP} - \parbox{10mm}{\includegraphics[scale=1.0]{fig10.eps}}_{(1)} \Biggl) +  \frac{N - 1}{9}\Biggl(\parbox{10mm}{\includegraphics[scale=1.0]{fig10.eps}}_{SP} - \parbox{10mm}{\includegraphics[scale=1.0]{fig10.eps}}_{(N - 1)} \Biggl) \Biggl] \Biggl\}, 
\end{eqnarray}
\begin{eqnarray}\label{uhdsuhdfuhdfu}
&& \Gamma_{T}(P^{2},t,M) = P^{2} + \frac{1}{6}R_{\mu\nu}\frac{\partial}{\partial P_{\mu}}\frac{P^{\nu}}{(P^{2} + t + g^{\ast}M^{2}/6)^{2}}  + t + \frac{1}{6}g^{\ast}M^{2} + \nonumber \\ && \frac{1}{6}g^{\ast}\Biggl\{ \Biggl[ \parbox{12mm}{\includegraphics[scale=1.0]{fig1.eps}}_{(1)} + (t + g^{\ast}M^{2}/2)\parbox{10mm}{\includegraphics[scale=1.0]{fig10.eps}}_{SP} \Biggl]  + (N + 1)\Biggl[ \parbox{12mm}{\includegraphics[scale=1.0]{fig1.eps}}_{(N - 1)} + (t + g^{\ast}M^{2}/6)\parbox{10mm}{\includegraphics[scale=1.0]{fig10.eps}}_{SP} \Biggl] + \nonumber \\ && \frac{2}{3}g^{\ast}M^{2}\Biggl[ \Biggl(\parbox{10mm}{\includegraphics[scale=1.0]{fig10.eps}}_{SP} - \parbox{10mm}{\includegraphics[scale=1.0]{fig10.eps}}_{(1,(N - 1))} \Biggl) \Biggl] \Biggl\}, 
\end{eqnarray}
where $M =\,<\phi>$, $g$ and $t$ are the renormalized magnetization (as a mean value of the renormalized field), dimensionless coupling constant and composite field coupling constant. The dimensionless and dimensionful coupling constants are related through $\lambda = g\kappa^{\epsilon/2}$, where $\kappa$ is some arbitrary momentum scale. The quantity $g^{\ast}$ is the dimensionless coupling constant evaluated at its nontrivial fixed point value. The nontrivial fixed point is computed from the nontrivial solution of the equation $\beta(g^{\ast}) = 0$ \cite{Amit}, where $\beta(g)$ is the $\beta$-function giving the flow of the renormalized dimensionless coupling constant flowing from some arbitrary value to the renormalized one when the renormalized theory is attained. This nontrivial fixed point is responsible for the computation of the loop quantum corrections to the amplitude ratios beyond the Landau approximation. The renormalized massless free propagator  $G_{0}(q) \equiv \parbox{12mm}{\includegraphics[scale=1.0]{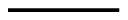}}$ is given by
\begin{eqnarray}
G_{0}(q) = \frac{1}{q^{2}} + \frac{(1/3 - \xi)R}{(q^{2})^{2}} - \frac{2R_{\mu\nu}q^{\mu}q^{\nu}}{3(q^{2})^{3}},
\end{eqnarray}
where we have to set $\xi \rightarrow \xi(d)$ at the middle of Feynman diagrams calculations to get rid the infrared divergences that yet would persist in the theory if we would maintain $\xi$ arbitrary and $\xi(d) = [(d - 2)/4(d - 1)]$ at arbitrary dimensions less than four \cite{HATHRELL1982136,PhysRevD.25.1019,Buchbinder.Odintsov.Shapiro,Parker.Toms,0264-9381-25-10-103001}, where $d = 4 - \epsilon$. . The subscript ``$SP$'' in the ``fish'' diagram $\parbox{8mm}{\includegraphics[scale=.8]{fig10.eps}}_{SP}$ means that this diagram is to be evaluated in the so called symmetry point, where the external momenta $P$ are held at fixed values according to $P_{i}\cdot P_{j} = (\kappa^{2}/4)(4\delta_{ij}-1)$, implying that $(P_{i} + P_{j})^{2} \equiv P^{2} = \kappa^{2}$ for $i\neq j$ \cite{Amit}. The nontrivial fixed point can be computed after we have evaluated the $\beta$-function displayed just below
\begin{eqnarray}
\beta(g) = -\epsilon g + \frac{N + 8}{6}\Bigg[ 1 + \frac{1}{2}\epsilon + \Bigg(\frac{R}{6\mu^{2}} - \frac{R_{\mu\nu}\widehat{P}^{\mu}\widehat{P}^{\nu}}{3\mu^{2}}\Bigg)\epsilon \Bigg]g^{2}  - \frac{3N + 14}{12}g^{3},
\end{eqnarray} 
where we have used the following evaluated Feynman diagrams
\begin{eqnarray}\label{khjhuh}
\parbox{10mm}{\includegraphics[scale=1.0]{fig10.eps}}_{SP} = \frac{1}{\epsilon} \Biggr(1 + \frac{1}{2}\epsilon + \frac{R}{6\mu^{2}}\epsilon - \frac{R_{\mu\nu}\widehat{P}^{\mu}\widehat{P}^{\nu}}{3\mu^{2}}\epsilon \Biggr),
\end{eqnarray}
\begin{eqnarray}\label{gfjfgujfguj}
\parbox{12mm}{\includegraphics[scale=0.8]{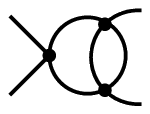}}_{SP} = \frac{1}{2\epsilon^{2}} \Biggr(1 + \frac{3}{2}\epsilon + \frac{R}{3\mu^{2}}\epsilon - 2\frac{R_{\mu\nu}P^{\mu}P^{\nu}}{3\mu^{4}}\epsilon \Biggr).
\end{eqnarray}
We emphasize that in the present method, the $\beta$-function depends on the nonuniversal curved spacetime parameters $R$ and $R_{\mu\nu}$ as well as the corresponding nontrivial fixed point
\begin{eqnarray}\label{klkoilkj}
g^{\ast} =  \frac{6\epsilon}{(N + 8)}\Bigg\{ 1 + \epsilon\left[ \frac{(9N + 42)}{(N + 8)^{2}} -\frac{1}{2} - \Bigg(\frac{R}{6\mu^{2}} - \frac{R_{\mu\nu}\widehat{P}^{\mu}\widehat{P}^{\nu}}{3\mu^{2}}\Bigg) \right]\Bigg\}.
\end{eqnarray}
Thus, following the notation of Ref. \cite{Neto2017} we obtain the critical amplitudes
\begin{eqnarray}\label{hduhsduhsdu}
&& A^{+} = \frac{N}{4}\Bigg[ 1 + \Bigg( \frac{4}{4 - N} + A_{N} \Bigg)\epsilon \Bigg], 
\end{eqnarray}
\begin{eqnarray}\label{jdoijdio}
A^{-} = 1 + \Bigg( \frac{N}{4 - N} - \frac{4 - N}{2(N + 8)}\ln 2 + A_{N} \Bigg)\epsilon ,
\end{eqnarray}     
where
\begin{eqnarray}\label{hduhsduhsduu}
&& A_{N} = \frac{1}{2} - \frac{9N + 42}{(N + 8)^{2}} - \frac{4 - N}{(N + 8)} - \frac{(N + 2)(N^{2} + 30N + 56)}{2(4 - N)(N + 8)^{2}} + \nonumber \\ && \frac{2(N + 2)}{N + 8}\Bigg(\frac{R}{6\mu^{2}} - \frac{R_{\mu\nu}\widehat{P}^{\mu}\widehat{P}^{\nu}}{3\mu^{2}}\Bigg), 
\end{eqnarray}
\begin{eqnarray}\label{jhdshudfhuifh}
C^{+} = 1 - \frac{N + 2}{2(N + 8)} \Bigg[ 1 + 2\Bigg(\frac{R}{6\mu^{2}} - \frac{R_{\mu\nu}\widehat{P}^{\mu}\widehat{P}^{\nu}}{3\mu^{2}}\Bigg) \Bigg]\epsilon , 
\end{eqnarray}
\begin{eqnarray}\label{ygdsuygsduygds}
C^{-} = \frac{1}{2}\Bigg\{1 - \frac{1}{6}\Bigg[ 4 + \ln 2 + 2\Bigg(\frac{R}{6\mu^{2}} - \frac{R_{\mu\nu}\widehat{P}^{\mu}\widehat{P}^{\nu}}{3\mu^{2}}\Bigg) \Bigg]\epsilon\Bigg\},
\end{eqnarray}
\begin{eqnarray}\label{fhduhdfu}
D = \frac{1}{6}g^{\ast(\delta - 1)/2}\Bigg\{1 + \Bigg[ 1 - \ln 2 - \frac{N - 1}{N + 8}\ln 3 +  2\Bigg(\frac{R}{6\mu^{2}} - \frac{R_{\mu\nu}\widehat{P}^{\mu}\widehat{P}^{\nu}}{3\mu^{2}}\Bigg) \Bigg]\epsilon\Bigg\},
\end{eqnarray} 
\begin{eqnarray}\label{huhfuhfuhf}
&& B = \Bigg(\frac{N + 8}{\epsilon}\Bigg\{1 -  \frac{3}{N + 8}\Bigg[ 1 + \ln 2 + 2\Bigg(\frac{R}{6\mu^{2}} - \frac{R_{\mu\nu}\widehat{P}^{\mu}\widehat{P}^{\nu}}{3\mu^{2}}\Bigg) \Bigg]\epsilon - \nonumber \\ && \Bigg[\frac{9N + 42}{(N + 8)^{2}} -\frac{1}{2} - \Bigg(\frac{R}{6\mu^{2}} - \frac{R_{\mu\nu}\widehat{P}^{\mu}\widehat{P}^{\nu}}{3\mu^{2}}\Bigg) \Bigg]\epsilon \Bigg\}\Bigg)^{1/2}, 
\end{eqnarray} 
\begin{eqnarray}\label{hjfdhjfhjf}
\xi_{0}^{+} = 1 - \frac{N + 2}{4(N + 8)} \Bigg[ 1 + 2\Bigg(\frac{R}{6\mu^{2}} - \frac{R_{\mu\nu}\widehat{P}^{\mu}\widehat{P}^{\nu}}{3\mu^{2}}\Bigg) \Bigg]\epsilon , 
\end{eqnarray}
\begin{eqnarray}\label{kjlkjhlkh}
\xi_{0}^{-} = 2^{-1/2}\Bigg\{1 - \frac{1}{12}\Bigg[ \frac{7}{2} +\ln 2 + 2\Bigg(\frac{R}{6\mu^{2}} - \frac{R_{\mu\nu}\widehat{P}^{\mu}\widehat{P}^{\nu}}{3\mu^{2}}\Bigg) \Bigg]\epsilon\Bigg\},
\end{eqnarray} 
\begin{eqnarray}\label{fdgfdsadfs}
&& \xi_{0}^{T} = \Biggl( \frac{\epsilon}{N + 8}\Bigg\{1 + \frac{3}{N + 8}\Bigg[ \frac{5}{6} + \ln 2 + 2\Bigg(\frac{R}{6\mu^{2}} - \frac{R_{\mu\nu}\widehat{P}^{\mu}\widehat{P}^{\nu}}{3\mu^{2}}\Bigg) \Bigg]\epsilon + \nonumber \\ && \Bigg[\frac{9N + 42}{(N + 8)^{2}} -\frac{1}{2} - \Bigg(\frac{R}{6\mu^{2}} - \frac{R_{\mu\nu}\widehat{P}^{\mu}\widehat{P}^{\nu}}{3\mu^{2}}\Bigg)\Bigg]\epsilon \Bigg\}\Biggl)^{1/(d - 2)},\nonumber \\  
\end{eqnarray} 
\begin{eqnarray}\label{ouypuopuo}
&& C^{c} = \frac{2D^{1/\delta}}{g^{\ast 1/2\beta}} \Bigg( 1 - \frac{9}{2(N + 8)}\Bigg\{ \Biggl[ 1 - \ln 2 + 2\Bigg(\frac{R}{6\mu^{2}} - \frac{R_{\mu\nu}\widehat{P}^{\mu}\widehat{P}^{\nu}}{3\mu^{2}}\Bigg) \Bigg] + \nonumber \\ &&  \frac{N - 1}{9}\Biggl[ 1 - \ln 6 + 2\Bigg(\frac{R}{6\mu^{2}} - \frac{R_{\mu\nu}\widehat{P}^{\mu}\widehat{P}^{\nu}}{3\mu^{2}}\Bigg) \Bigg] + \frac{2(N + 8)}{27} \Bigg\}\epsilon\Bigg),
\end{eqnarray}
\begin{eqnarray}\label{bxvbcvbxvc}
&& \xi_{0}^{c} = \frac{2^{1/2}D^{1/2\delta}}{g^{\ast 1/4\beta}} \Bigg( 1 - \frac{9}{4(N + 8)}\Bigg\{ \Biggl[ 1 - \ln 2 + 2\Bigg(\frac{R}{6\mu^{2}} - \frac{R_{\mu\nu}\widehat{P}^{\mu}\widehat{P}^{\nu}}{3\mu^{2}}\Bigg) \Bigg] + \nonumber \\ &&  \frac{N - 1}{9}\Biggl[ 1 - \ln 6 + 2\Bigg(\frac{R}{6\mu^{2}} - \frac{R_{\mu\nu}\widehat{P}^{\mu}\widehat{P}^{\nu}}{3\mu^{2}}\Bigg) \Bigg] + \frac{N + 14}{27} \Bigg\}\epsilon\Bigg)^{1/2},
\end{eqnarray}
\begin{eqnarray}\label{hjdghssf}
 \widehat{D} = 1,
\end{eqnarray}
where the remaining evaluated Feynman diagrams needed to obtain the critical amplitudes aforementioned are the ones shown in the Eq. (\ref{khjhuh}) and in the ones (\ref{fldsjlkfdjlk})-(\ref{tyrttffyty}) 
\begin{eqnarray}\label{fldsjlkfdjlk}
\parbox{12mm}{\includegraphics[scale=1.0]{fig1.eps}}_{(1)} = -\frac{t + g^{\ast}M^{2}/2}{\epsilon}\Biggl[ 1 - \frac{1}{2}\ln \Biggl( t + \frac{g^{\ast}M^{2}}{2} \Biggl)\epsilon \Biggl],  
\end{eqnarray}
\begin{eqnarray}\label{fldsjlkfdjl}
&& \parbox{11mm}{\includegraphics[scale=1.0]{fig1.eps}}_{(N - 1)} = -\frac{t + g^{\ast}M^{2}/6}{\epsilon}\Biggl[ 1 - \frac{1}{2}\ln \Biggl( t + \frac{g^{\ast}M^{2}}{6} \Biggl)\epsilon \Biggl],  
\end{eqnarray}
\begin{eqnarray}
\parbox{10mm}{\includegraphics[scale=1.0]{fig10.eps}}_{(1)} =  \frac{1}{\epsilon} \Biggr[1 - \frac{1}{2}\epsilon - \frac{1}{2}\epsilon L_{1}(P^{2})  + \frac{R}{6\mu^{2}}\epsilon L_{R,1}(P^{2}) - \frac{R_{\mu\nu}P^{\mu}P^{\nu}}{3\mu^{4}}\epsilon L_{R_{\mu\nu},1}(P^{2}) \Biggr],
\end{eqnarray}
\begin{eqnarray}
&& \parbox{10mm}{\includegraphics[scale=1.0]{fig10.eps}}_{(N - 1)} = \nonumber \\ && \frac{1}{\epsilon} \Biggr[1 - \frac{1}{2}\epsilon - \frac{1}{2}\epsilon L_{(N-1)}(P^{2})   + \frac{R}{6\mu^{2}}\epsilon L_{R,(N - 1)}(P^{2}) - \frac{R_{\mu\nu}P^{\mu}P^{\nu}}{3\mu^{4}}\epsilon L_{R_{\mu\nu},(N - 1)}(P^{2}) \Biggr], 
\end{eqnarray}
\begin{eqnarray}\label{assdxe}
&& \parbox{10mm}{\includegraphics[scale=1.0]{fig10.eps}}_{1,(N - 1)} = \nonumber \\ && \frac{1}{\epsilon} \Biggr[1 - \frac{1}{2}\epsilon - \frac{1}{2}\epsilon L_{1,(N-1)}(P^{2}) +  \frac{R}{6\mu^{2}}\epsilon L_{R,[1,(N - 1)]}(P^{2}) -  \frac{R_{\mu\nu}P^{\mu}P^{\nu}}{3\mu^{4}}\epsilon L_{R_{\mu\nu},[1,(N - 1)]}(P^{2}) \Biggr],
\end{eqnarray}
\begin{eqnarray}\label{et5rtgdgff}
L_{1}(P^{2}) = \int_{0}^{1}dx \ln \left[\frac{x(1-x)P^{2} + t + \frac{g^{\ast}M^{2}}{2}}{\mu^{2}}\right],
\end{eqnarray}
\begin{eqnarray}\label{jhkjoji}
L_{(N-1)}(P^{2}) = \int_{0}^{1}dx \ln \left[\frac{x(1-x)P^{2} + t + \frac{g^{\ast}M^{2}}{6}}{\mu^{2}}\right],
\end{eqnarray}
\begin{eqnarray}\label{gjfghdfyr}
L_{1,(N-1)}(P^{2}) =  \int_{0}^{1}dx \ln \left[\frac{x(1-x)P^{2} + x(t + \frac{g^{\ast}M^{2}}{2}) + (1-x)(t + \frac{g^{\ast}M^{2}}{6})}{\mu^{2}}\right],\nonumber \\   
\end{eqnarray}
\begin{eqnarray}\label{zcxvcgfd}
L_{R,1}(P^{2}) =  \int_{0}^{1}d x\frac{x(1 - x)}{\frac{x(1 - x)P^{2}}{\mu^{2}} + \frac{t + \frac{g^{\ast}M^{2}}{2}}{\mu^{2}}},
\end{eqnarray}
\begin{eqnarray}\label{jlkjiyiuj}
L_{R_{\mu\nu},1}(P^{2}) =  \int_{0}^{1}d x\frac{x^{2}(1 - x)^{2}}{\left[\frac{x(1 - x)P^{2}}{\mu^{2}} + \frac{t + \frac{g^{\ast}M^{2}}{2}}{\mu^{2}}\right]^{2}},
\end{eqnarray}
\begin{eqnarray}\label{bmbnbmv}
L_{R,(N - 1)}(P^{2}) =  \int_{0}^{1}d x\frac{x(1 - x)}{\frac{x(1 - x)P^{2}}{\mu^{2}} + \frac{t + \frac{g^{\ast}M^{2}}{6}}{\mu^{2}}},
\end{eqnarray}
\begin{eqnarray}\label{wereretjjgdhg}
L_{R_{\mu\nu},(N - 1)}(P^{2}) =  \int_{0}^{1}d x\frac{x^{2}(1 - x)^{2}}{\left[\frac{x(1 - x)P^{2}}{\mu^{2}} + \frac{t + \frac{g^{\ast}M^{2}}{6}}{\mu^{2}}\right]^{2}},
\end{eqnarray}
\begin{eqnarray}\label{ipkokljgdhg}
L_{R,[1,(N - 1)]}(P^{2}) =  \int_{0}^{1}d x\frac{x(1 - x)}{\frac{x(1 - x)P^{2}}{\mu^{2}} + \frac{x(t + \frac{g^{\ast}M^{2}}{2}) + (1-x)(t + \frac{g^{\ast}M^{2}}{6})}{\mu^{2}}},
\end{eqnarray}
\begin{eqnarray}\label{tyrttffyty}
L_{R_{\mu\nu},[1,(N - 1)]}(P^{2}) = \int_{0}^{1}d x\frac{x^{2}(1 - x)^{2}}{\left[\frac{x(1 - x)P^{2}}{\mu^{2}} + \frac{x(t + \frac{g^{\ast}M^{2}}{2}) + (1-x)(t + \frac{g^{\ast}M^{2}}{6})}{\mu^{2}}\right]^{2}},\quad\quad
\end{eqnarray}
where $\widehat{P}^{\mu}$ is dimensionless and unitary such that $P^{\mu} = \kappa \widehat{P}^{\mu}$ \cite{EurophysLett10821001}. Thus as for the $\beta$-function and nontrivial fixed point, the amplitudes themselves depend on the nonuniversal curved spacetime parameters $R$ and $R_{\mu\nu}$. But now, if we apply the expressions above for the amplitudes, we have the cancelling of the curved spacetime parameters $R$ and $R_{\mu\nu}$ in the amplitude ratios computation and obtain that the curved spacetime amplitude ratios are identical to their flat spacetime counterparts \cite{V.PrivmanP.C.Hohenberg}. These results show that the amplitude ratios values are insensible to the conformal symmetry thus maintaining the two-scale-factor universality hypothesis validity intact. In fact, as the symmetry under consideration is one present in the space where the quantum field is embedded and not in its internal one, the curved spacetime amplitude ratios values must be the same as that of flat sapetime ones. Just a change in the internal symmetry of the field could modify the amplitude ratios values.

\section{Minimal subtraction scheme}\label{Minimal subtraction scheme}

\par This present method is much more general and elegant than the earlier \cite{Amit}, since in the normalization conditions method, the external momenta are fixed at some value, namely the symmetry point while in the present one they are kept at general arbitrary values. In the present method we consider only the divergent part of the diagrams and not both divergent and finite ones as in the earlier method. Thus, for example, instead of considering the $\parbox{8mm}{\includegraphics[scale=.8]{fig10.eps}}_{SP}$ diagram in the Eqs. (\ref{fhdldglkjdflk})-(\ref{uhdsuhdfuhdfu}) we have to consider the $[\parbox{8mm}{\includegraphics[scale=.8]{fig10.eps}}]_{S}$ one where $[\parbox{8mm}{\includegraphics[scale=.8]{fig10.eps}}]$ is that of Eq. (\ref{klijouio}) and $[~~~]_{S}$ means that the only part of the diagram we have to consider is its singular and one so on for the remaining diagrams. Thus, in this method, we obtain amplitudes which are different from that obtained through the earlier method. Their values in this method are given by
\begin{eqnarray}\label{terreteu}
&& A^{+} = \frac{N}{4}\Bigg[ 1 + \Bigg( \frac{4}{4 - N} + A_{N}^{\prime} \Bigg)\epsilon \Bigg], 
\end{eqnarray}
\begin{eqnarray}\label{eytwertyer}
 A^{-} = \Bigg[1 + \Bigg( \frac{N}{4 - N} - \frac{4 - N}{2(N + 8)}\ln 2 + A_{N}^{\prime} \Bigg)\epsilon\Bigg],
\end{eqnarray}     
where
\begin{eqnarray}\label{reytreytrwe}
A_{N}^{\prime} = - \frac{9N + 42}{(N + 8)^{2}} - \frac{4 - N}{(N + 8)} - \frac{(N + 2)(N^{2} + 30N + 56)}{2(4 - N)(N + 8)^{2}},
\end{eqnarray}
\begin{eqnarray}\label{uiyiupoiu}
&& C^{+} = 1, 
\end{eqnarray}
\begin{eqnarray}\label{iuoipiyo}
&& C^{-} = \frac{1}{2}\Bigg[1 - \frac{1}{6}( 3 + \ln 2 )\epsilon\Bigg],
\end{eqnarray}  
\begin{eqnarray}\label{tretrety}
D = \frac{1}{6}g^{\ast(\delta - 1)/2}\Bigg[1 - \frac{1}{2}\Bigg( \ln 2 + \frac{N - 1}{N + 8}\ln 3 \Bigg)\epsilon\Bigg],
\end{eqnarray} 
\begin{eqnarray}\label{tuytyut}
B = \Bigg\{\frac{N + 8}{\epsilon}\Bigg[1 - \frac{3\ln 2}{N + 8}\epsilon - \frac{9N + 42}{(N + 8)^{2}}\epsilon \Bigg]\Bigg\}^{1/2},
\end{eqnarray} 
\begin{eqnarray}\label{xcvbcxzvcxzv}
\xi_{0}^{+} = 1,
\end{eqnarray}
\begin{eqnarray}\label{xvnbxvznbx}
\xi_{0}^{-} = 2^{-1/2}\Bigg[1 - \frac{1}{12}\Bigg( \frac{5}{2} +\ln 2 \Bigg)\epsilon\Bigg],
\end{eqnarray}
\begin{eqnarray}\label{tryteyt}
\xi_{0}^{T} =  \Biggl\{ \frac{\epsilon}{(N + 8)}\Bigg[1 + \frac{3}{N + 8}\Bigg( -\frac{1}{6} + \ln 2 \Bigg)\epsilon + \frac{9N + 42}{(N + 8)^{2}}\epsilon \Bigg]\Biggl\}^{1/(d - 2)},
\end{eqnarray} 
\begin{eqnarray}\label{bcnbvcbc}
C^{c} = \frac{2D^{1/\delta}}{g^{\ast 1/2\beta}} \Bigg\{ 1 - \frac{9}{2(N + 8)}\Bigg[ - \ln 2 - \frac{N - 1}{9}\ln 6 + \frac{2(N + 8)}{27} \Bigg]\epsilon\Bigg\},
\end{eqnarray}
\begin{eqnarray}\label{bvnbcvb}
\xi_{0}^{c} = \frac{2^{1/2}D^{1/2\delta}}{g^{\ast 1/4\beta}} \Bigg\{ 1 - \frac{9}{4(N + 8)}\Bigg[ - \ln 2 - \frac{N - 1}{9}\ln 6 + \frac{N + 14}{27} \Bigg]\epsilon\Bigg\}^{1/2},
\end{eqnarray}
\begin{eqnarray}\label{eetetee}
\widehat{D} = 1, 
\end{eqnarray}
where the nontrivial fixed point is obtained through the corresponding $\beta$-function
\begin{eqnarray}
\beta(g) = -\epsilon g + \frac{N + 8}{6}g^{2} - \frac{3N + 14}{12}g^{3},
\end{eqnarray} 
computed from the evaluated Feynman diagrams (\ref{klijouio})-(\ref{bnnmbj}) just displayed below
\begin{eqnarray}\label{klijouio}
\parbox{10mm}{\includegraphics[scale=1.0]{fig10.eps}} = \frac{1}{\epsilon} \Biggr[1 - \frac{1}{2}\epsilon - \frac{1}{2}\epsilon J(P^{2})  + \frac{R}{6\mu^{2}}\epsilon J_{R}(P^{2}) - \frac{R_{\mu\nu}P^{\mu}P^{\nu}}{3\mu^{4}}\epsilon J_{R_{\mu\nu}}(P^{2}) \Biggr],
\end{eqnarray}
\begin{eqnarray}\label{bnnmbj}
\parbox{12mm}{\includegraphics[scale=0.8]{fig21.eps}} = \frac{1}{2\epsilon^{2}} \Biggr[1 - \frac{1}{2}\epsilon - \epsilon J(P^{2})  + \frac{R}{3\mu^{2}}\epsilon J_{R}(P^{2}) -  2\frac{R_{\mu\nu}P^{\mu}P^{\nu}}{3\mu^{4}}\epsilon J_{R_{\mu\nu}}(P^{2})\Biggr],
\end{eqnarray}
where 
\begin{eqnarray}\label{ugujdfjgdhg}
J_{R}(P^{2}) =  \int_{0}^{1}d x\frac{x(1 - x)}{x(1 - x)P^{2}},
\end{eqnarray}
\begin{eqnarray}\label{ufgfghujjgdhg}
J_{R_{\mu\nu}}(P^{2}) =  \int_{0}^{1}d x\frac{x^{2}(1 - x)^{2}}{\left[x(1 - x)P^{2}\right]^{2}}.
\end{eqnarray}
The nontrivial fixed point s given by  
\begin{eqnarray}\label{klkoilkj}
g^{\ast} = \frac{6\epsilon}{(N + 8)}\Bigg\{ 1 + \epsilon\left[ \frac{(9N + 42)}{(N + 8)^{2}} \right]\Bigg\}.
\end{eqnarray} 
As this method is more general and elegant, the $R$ and $R_{\mu\nu}$ terms dependent on the curved spacetime cancel out already in the amplitude ratios, $\beta$-function and nontrivial fixed point computation, such that the corresponding curved spacetime quantities themselves are the same as that of flat spacetime. Thus, the resulting curved spacetime amplitude ratios are automatically the same as their flat spacetime counterparts.

\section{Massless BPHZ method}\label{Massless BPHZ method}
   
\par The massless BPHZ method \cite{BogoliubovParasyuk,Hepp,Zimmermann} is different from the two earlier ones. While in the two former ones the divergences of the theory are absorbed by a procedure starting from the bare theory, in the former one the divergences are absorbed starting from the renormalized theory by introducing counterterms diagrams. Then only the divergent part of the resulting diagrams are to be considered and the operator which represents this operation is given by $\mathcal{K}(~~~)$ \cite{Neto2017} and acts in the corresponding diagrams. Thus, although the  present method is distinct from that of earlier section, at least at the loop level considered here, they lead to the same results. Now we have to present our conclusions. 

\section{Conclusions and perspectives}

\par In this Letter we have probed the effect of conformal symmetry on the curved spacetime amplitude ratios for massless O($N$) $\lambda\phi^{4}$ scalar field theories in curved spacetime. By applying three distinct and independent field-theoretic renormalization group methods, the amplitude ratios have been evaluated in the normalization conditions, minimal subtraction scheme and massless BPHZ methods, respectively, where the external momenta of Feynman diagrams have been held at fixed values in the former method and at arbitrary values in the later ones. We have found, at least at the loop level considered, identical curved spacetime amplitude ratios values when obtained through the three methods. We emphasize the importance of employing more than one method for computing the amplitude ratios since we can check the final results when obtained through that methods. Furthermore, the curved spacetime amplitude ratios obtained having been the same as their flat spacetime counterparts. This fact has confirmed the two-scale-factor universality hypothesis validity since the symmetry probed here has been one defined in the spacetime where the field is embedded and not in its internal one. We believe that the influence of conformal symmetry on the critical properties of systems undergoing continuous phase transitions can be analogously probed in future works through the evaluation of finite-size scaling effects, corrections to scaling etc.       

\section*{Acknowledgements}

\par HASC and PRSC would like to thank CAPES (brazilian funding agency) for financial support. Furthermore, PRSC would like to thank CNPq (brazilian funding agency) for financial support through Grant number:  431727/2018.

\bibliography{apstemplate}

\end{document}